\documentclass{jetpl}
\twocolumn
\lat

\title{Comment on ``Order parameter of $A$-like $^3$He phase in aerogel''}
\rtitle{Comment on ``Order parameter of $A$-like $^3$He phase in aerogel''}
\sodtitle{Comment on ``Order parameter of $A$-like $^3$He phase in aerogel''}

\author{V.\,P.\,Mineev\/\thanks{mineev@drfmc.ceng.cea.fr}
and M.\,E.\,Zhitomirsky}
\rauthor{V.\,P.\,Mineev, M.\,E.\,Zhitomirsky}
\sodauthor{V.\,P.\,Mineev, M.\,E.\,Zhitomirsky}

\address{Commissariat \'a l'Energie Atomique, DSM/DRFMC/SPSMS 
38054 Grenoble, France}
\dates{\today}{*}   

\abstract{We argue that the inhomogeneous $A$-phase in
aerogel is energetically more preferable than the ``robust'' phase
suggested by I. A. Fomin, 
JETP Lett. {\bf 77}, 240 (2003); cond-mat/0302117/0401639.}
\begin{document}

\maketitle

Experimental investigation of the superfluid phases of $^3$He in aerogel
is at present a hot subject in low temperature physics (see the most recent
publications \cite{Baumgardner,Choi} and references
therein). In view of its anisotropic properties, a special interest
has been attracted to the $A$-like superfluid phase.  
As was pointed out by Volovik \cite{Volovik} such a   
phase corresponds at short length scale to the ordinary $A$-phase,
while at larger distances it presents a kind of
superfluid glass with irregular distribution of the
direction of Cooper pairs angular momentum and absence of superfluid
properties.  Volovik's derivation has been based on the general analysis
due to Imry and Ma of phase transitions with breaking of a  
continuous symmetry
in the presence of random local anisotropy
\cite{Imry}.  
Recently, Fomin has published a series of papers
\cite{Fomin2,Fomin3,Fomin4} where he claims that ``general
argument of Imry and Ma does not directly apply to the superfluid
$^3$He in aerogel.''  He has introduced  anisotropic interaction of
the superfluid $^3$He with aerogel
\begin{equation}
F_\eta = \int \eta_{ij}^{(a)}({\bf r})A_{\mu i}({\bf r})
A^*_{\mu j}({\bf r})\,d^3 r \ ,
\label{1}
\end{equation}
where $A_{\mu i}$ is the superfluid order parameter
and a traceless position-dependent tensor
$\eta_{ij}^{(a)}=\eta_{ij}-\frac{1}{3}\eta_{ll}\delta_{ij}$
describes local splitting of $T_c$ for different projections of
angular momenta because of anisotropic suppression of superfluidity by
aerogel strands.  Isotropic part of this tensor subtracted here is
included in a term, which produces a local shift of the critical temperature.
Due to the time reversal invariance of the energy $F_\eta$
the tensor $\eta_{ij}^{(a)}$ obeys the symmetry
$\eta_{ij}^{(a)}=\eta_{ji}^{(a)}$.

According to Fomin \cite{Fomin4} the interaction (\ref{1})
plays the role of the ``surface'' energy,
which is lost for any superfluid phase except for the case when
there is an average value of the order parameter $\bar{A}_{\mu i}$
such that
\begin{equation}
\eta_{ij}^{(a)}\bar{A}_{\mu i}\bar{A}^*_{\mu j} =0
\end{equation}
or, equivalently,
\begin{equation}
{\bar A_{\mu i}}{\bar A^*_{\mu j}}+ {\bar A_{\mu j}¥}{\bar
A^*_{\mu i}} \propto \delta_{ij} \ .
\label{2}
\end{equation}
The above constraint removes the ``surface'' term $F_\eta\equiv 0$
and leads to the 
conclusion \cite{Fomin4} that superfluid phases of $^3$He in
aerogel below the second order transition from the normal state should
satisfy Eq.~(\ref{2}).  The $B$-phase with
$A^B_{\mu i}=\Delta_B R_{\mu i}e^{i\varphi}$ does satisfy this
condition, but for the ordinary $A$-phase Eq.~(\ref{2}) is not fulfilled.
The $A$-phase order parameter is given by
\begin{equation}
A^A_{\mu i}=\Delta_A V_{\mu}(m_i+in_i) \ ,
\label{3}
\end{equation}
where a unit vector $V_{\mu}$ determines orientation 
of the spin quantization axis, while two orthogonal vectors
${\bf m}$ and ${\bf n}$ yield 
direction of the orbital momentum ${\bf l}={\bf m}\times{\bf n}$.  
As a result, it has been proposed
to consider instead of the $A$-phase a
class of so called ``robust'' phases satisfying Eq.~(\ref{2})
\cite{Fomin2,Fomin3,Fomin4}.

Let us, nevertheless, substitute the
$A$-phase order parameter into Eq.~(\ref{1}):
\begin{equation}
F_{\eta}=\Delta^2_A \int \eta_{ij}^{(a)}[m_i({\bf r})m_j({\bf r})+
n_i({\bf r})n_j({\bf r})]\,d^3r \ .
\label{4}
\end{equation}
Using identity $m_i({\bf r})m_j({\bf r})+
n_i({\bf r})n_j({\bf r})+l_i({\bf r})l_j({\bf r})
=\delta_{ij}$ we obtain
\begin{equation}
F_\eta= -\Delta^2_A \int\eta_{ij}^{(a)}l_i({\bf r})l_j({\bf r})\,
d^3r \ .
\label{5}
\end{equation}
Any uniform state of the $A$-phase has $F_\eta=0$, 
since $\int\eta_{ij}\,d^3r=0$. This is, actually, true for an arbitrary
homogeneous phase, which is effectively ``robust'' on 
average and has the same transition temperature as the states (\ref{2}).  
The ``non-robust'' $A$-phase can further gain in energy from long-scale
fluctuations of the random anisotropy by adjusting the direction 
of vector ${\bf l}$ on a certain length-scale $L$.  
So, we just return to the standard Imry-Ma picture 
described in application to the superfluid $^3$He
by Volovik \cite{Volovik}. The only difference with the Imry-Ma
scenario is that space variations of the vector ${\bf l}({\bf r})$
do not destroy the phase transition: the complex superfluid
order parameter $A_{\mu i}({\bf r})$ breaks additional 
spin-rotational symmetry and partly the gauge symmetry \cite{Volovik}. 
Thus, the adjustment of vector $\bf l$ to the long-scale
fluctuations of the anisotropic energy leads to an {\it enhancement}
of the transition temperature of the generalized $A$-phase compared
to the critical temperature of 
the ``robust'' axi-planar state suggested by Fomin.  
The proper estimate of the domain-size $L$ can be found in \cite{Volovik}. 

As for the superfluid properties, the randomness of 
the distribution of ${\bf l}({\bf r})$
vector does not destroy the superfluid flow in $^3$He-$A$
in aerogel. There is, in fact, just the opposite effect:
fixing of the $\bf l$ direction prevents the phase slippage
processes and makes the $A$-phase in aerogel even a better
superfluid than in the bulk.

In conclusion, 
there is no reason for the stability of the ``robust''
phases, which have a higher energy than the locally homogeneous (on
length scale $L$) $A$-phase.


\begin{references}

\bibitem{Baumgardner} 
J.E. Baumgardner and D.D. Osheroff, Phys. Rev. Lett.
{\bf 93}, 155301 (2004).

\bibitem{Choi} 
H.C. Choi, A.J. Gray, C.L. Vincente, J.S. Xia, G. Gervais,
W.P. Halperin, N. Mulders, and Y. Lee, Phys. Rev. Lett.
{\bf 93}, 145302 (2004).

\bibitem{Volovik}
G.E. Volovik, Pis'ma Zh.  Eksp.  Teor.  Fiz.  {\bf 63},
281 (1996) [JETP Lett. {\bf 63}, 301 (1996)].

\bibitem{Imry}
Y. Imry and S. Ma, Phys. Rev. Lett. {\bf 35}, 1399 (1975).

\bibitem{Fomin2}
I.A. Fomin, Pis'ma Zh.  Eksp.  Teor.  Fiz.  {\bf 77},
285 (2003) [JETP Lett. {\bf 77}, 240 (2003)];
{\tt cond-mat/0302117}.

\bibitem{Fomin3}
I.A. Fomin, J. Low Temp. Phys.  {\bf 134}, 769 (2004).

\bibitem{Fomin4} 
I.A. Fomin, Zh. Eksp. Teor.  Fiz.  {\bf 125}, 1115 (2004)
[JETP, {\bf 98}, 974 (2004)] and {\tt cond-mat/0401639}.


\end{references}
\end{document}